\documentstyle[psfig,subfigure,astrobib]{mn-ab}
\title[Unlocking the Keyhole]
{Unlocking the Keyhole -- H$_2$ and PAH emission from molecular
clumps in the Keyhole Nebula}
\author[Brooks, Burton, Rathborne, Ashley, \& Storey]
{K. J.~Brooks\thanks{Current institute: European Southern Observatory,
Casilla 19001, Santiago 19, Chile (kbrooks@eso.org)}, M. G. Burton, J. M. Rathborne,  
\newauthor M. C. B. Ashley, J. W. V. Storey \\
School of Physics, The University of New South Wales, Sydney, NSW,
2052, Australia}

\newcommand{\kms}{km\,s$^{-1}$}
\newcommand{\Msun}{M$_{\odot}$}
\newcommand{\Lsun}{L$_{\odot}$}
\newcommand{\HII}{$\mathrm{H\,{\scriptstyle II}}\,$}
\newcommand{\CO}{$^{12}${\rmfamily CO}{(2--1)}$\,$}

\begin{document}
\maketitle
 
\begin{abstract}
To better understand the environment surrounding CO emission clumps in the
Keyhole Nebula, we have made images of the region in H$_2$ 1--0 S(1) (2.122
$\mu$m) emission and polycyclic aromatic hydrocarbon (PAH) emission at 3.29
$\mu$m. Our results show that the H$_2$ and PAH emission regions are
morphologically similar, existing as several clumps, all of which
correspond to CO emission clumps and dark optical features. The emission
confirms the existence of photodissociation regions (PDRs) on the surface
of the clumps. By comparing the velocity range of the CO emission with the
optical appearance of the H$_2$ and PAH emission, we present a model of the
Keyhole Nebula whereby the most negative velocity clumps are in front of
the ionization region, the clumps at intermediate velocities are in it and
those which have the least negative velocities are at the far side. It may
be that these clumps, which appear to have been swept up from molecular gas
by the stellar winds from $\eta$ Car, are now being over-run by the
ionization region and forming PDRs on their surfaces. These clumps comprise
the last remnants of the ambient molecular cloud around \mbox{$\eta$ Car}.

\end{abstract}
 
 
\section{Introduction}

\begin{figure*}
\centering \mbox{\subfigure[Three-colour optical emission image, courtesy
of David Malin.  This image was produced by combining 3 exposures (of 15
mins) each taken using blue, green and red filters. The Keyhole is the
prominent dark band bisecting the
image.]{\psfig{file=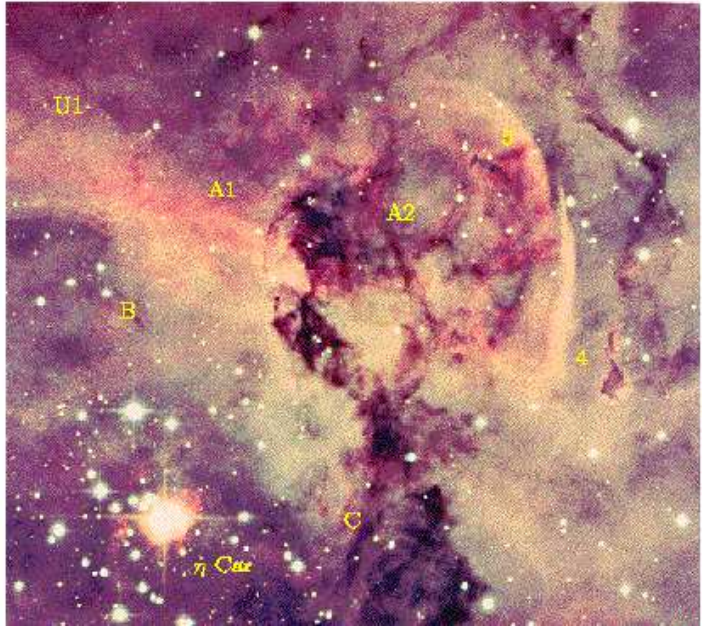,width=0.56\textwidth}}\quad}
\mbox{\subfigure[H$_2$  2.122 $\mu$m line emission obtained with
UNSWIRF. This image is continuum subtracted (aside from the circular
residuals from $\eta$ Car).]{\psfig{file=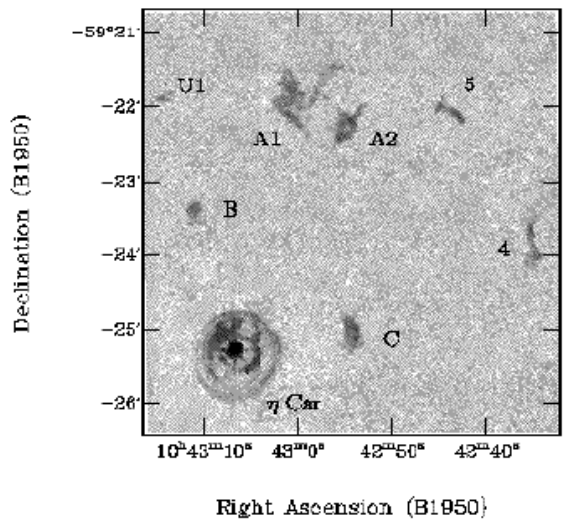,width=0.48\textwidth}}\quad \subfigure[PAH 3.29 $\mu$m emission obtained
with SPIREX. This image includes continuum sources (the diffraction spikes
from $\eta$ Car are artifacts).]{\psfig{bbllx=71pt,bblly=163pt,bburx=361pt,bbury=437pt,file=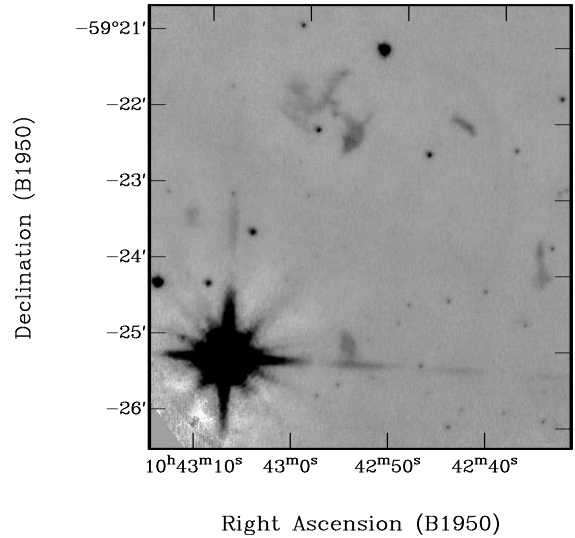,clip=,width=0.48\textwidth}}}
\caption{\label{multiwavelength}A collection of multi-wavelength images of
the Keyhole Nebula. Features of interest are labelled in the optical and
H$_2$ 1--0 S(1) images and are described further in Tables~\ref{co-clumps} and ~\ref{h2pah-table}.}
\end{figure*}

The Keyhole Nebula is part of the Carina Nebula, a star forming/giant
molecular cloud (GMC) complex, containing some of the most massive stars
known in our galaxy. It is bathed in the UV radiation field from the nearby
open star cluster, Tr 16. This cluster contains numerous O-type stars,
including three O3 stars \cite{Walborn95}. It is at a distance of 2.2 kpc
\cite{Tovmassian95} and is centred on one of the most massive stars known
-- $\eta$ Car, which has a present day mass of about 120 \Msun\
\cite{Davidson97}. The optical image of the Keyhole Nebula taken by
\citeN{Malin93} reveals many interesting features such as dark patches and
filaments, the most prominent being in the shape of an old-fashioned key
hole, which gave the nebula its name (see
Fig.~\ref{multiwavelength}a). \citeN{Herschel47} was the first to note
this peculiar feature and likened it to the shape of a
$``$lemniscata". Some of the smaller features also have bright rims. This
is beautifully illustrated in the recent colour-composite image taken with
NASA's Hubble Space Telescope.\footnote{These data were collected by the
Hubble Heritage Team, N. R. Walborn, R. H. Barba and A. Caulet and were
released  after the submission date of this article (see \\ http://oposite.stsci.edu/pubinfo/pr/2000/06/pr-photos.html).}

The molecular gas in this region is highly inhomogeneous, breaking into
several clumps which are all associated with the dark optical features
\cite{Cox951}. The clumps, typically a few tenths of a pc across
(10--30 arcsec) and with masses of $\sim 10$ \Msun\, are distributed
across a range of velocities (see Table~\ref{co-clumps}). Some of the
molecular clumps correspond to those optical features that have bright
rims, suggesting the presence of photodissociation regions (PDRs) at the
interface with the surrounding ionized gas. PDRs are regions where the
far-UV radiation field dominates the heating and chemistry of the gas. They
begin at the sharp ionized/neutral interface at the edge of molecular
clouds and then extend well into the molecular gas (see
\citeNP{Tielens93}).

To better understand the environment surrounding these clumps we have
imaged the Keyhole region at two wavelengths suitable for detecting
emission from PDRs. We have used the University of New South Wales Infrared
Fabry-Perot (UNSWIRF) to image H$_2$ 1--0 S(1) (2.122 $\mu$m) emission and
the SPIREX/Abu thermal infrared camera at the South Pole to image
narrow-band emission at 3.29 $\mu$m. This wavelength includes one of the
more prominent `unidentified' infrared (UIR) emission bands and is
generally attributed to the \mbox{C--H} stretching mode of free polycyclic
aromatic hydrocarbon molecules (PAHs) (\citeNP{Leger84},
\citeNP{Allamandola85}, \citeNP{Geballe94}, \citeNP{Joblin95}).

H$_2$ 1--0 S(1) emission in molecular clouds can be attributed to
collisional excitation in the presence of shocks or to UV fluorescence in
PDRs (see \citeNP{Burton92}). In dense gas, $ n \geq n_{crit} \sim 10^5$
cm$^{-3}$ where $n$ is the hydrogen nucleus density, UV excitation levels
are collisionally redistributed leading to an increase in the population of
the $\nu=1$ level. This subsequently increases the intensity of the
$\nu$=1--0 S(1) line (see \citeNP{Allen99}). Previous observations of PDRs
indicate that 3.29 $\mu$m PAH emission zones are situated near the very
edge of molecular clouds, peaking at the ionization front, while H$_2$ 1--0
S(1) emission zones extend further into the clouds by about 2 magnitudes of
visible extinction (e.g. \citeNP{Sloan97}, \citeNP{Tielens93},
\citeNP{Burton89}). So while H$_2$ 1--0 S(1) emission cannot be used alone
to identify PDRs, if found to correspond to another PDR tracer (such as PAH
emission) it can pinpoint regions of molecular gas that are high in density
and exposed to the surrounding far-UV radiation field.

\begin{table}
\begin{minipage}{\columnwidth}
\caption{\label{co-clumps}A list of the $^{12}$CO(2--1) emission clumps 
measured by Cox \& Bronfman (1995). The mass measurements may underestimate 
the true mass of each clump by up to a factor of 3. The type of dark optical 
feature each clump is associated with is also listed. The velocities are given
with respect to the local standard of rest (LSR).}
\begin{tabular}{@{}llll}
 
\hline
Clump	  	& Velocity 	&Mass 	 	&Optical\\
        	&(\kms)   	&(M$_{\sun}$) 	&Feature\\
\hline

C 	&$-7.6$		&6	&  Faint\\
B 	&$-8.5$		&3	&  Faint	\\
A1\footnote{Also known as the `Kangaroo Nebula'.}  &$-17.4$  &14  & Bright rim      \\
A2 	& --		& --	& Bright rim	\\
4 	&$-25.0$		&6	& Bright rim	\\
5 	&$-25.1$		&1	& Bright rim	\\
3 	&$-28.3$		&4 	& Keyhole (dark) \\
1 	&$-31.9$		&17 	& Keyhole (dark) \\
2 	&$-32.0$		&11	& Keyhole (dark) \\
\hline

\end{tabular} 
\end{minipage}
\end{table}

\section{Observations}
\subsection{UNSWIRF observations}
\label{unswirf-obs}
UNSWIRF \cite{Ryder98} is a near-infrared tunable imaging spectrometer. It
was used in conjunction with the Infrared Imager and spectrometer (IRIS) on
the Anglo-Australian Telescope (AAT), and produces a circular field of 1.7
arcmin diameter at 0.77 arcsec pixel$^{-1}$.

Images of H$_2$ 1--0 S(1) (2.122 $\mu$m) emission towards the Keyhole
region were obtained on 1996 April 24. A series of 30 pointings were made
on a 1 arcmin grid at the wavelength of the 2.122 $\mu$m line. An
integration time of 60 s was used. An image of the sky was obtained after
every five minutes, offset by 5 arcmin from the previous position. This
entire sequence was repeated at 3--4 profile widths from the line centre in
order to sample the corresponding continuum emission.

Bias subtraction and linearisation were performed during readout. All the
object images were sky subtracted and then flat-fielded using normalised
dome flat-fields. The continuum images, appropriately scaled, were then
subtracted to leave just pure line emission in each. However, the strong
emission from $\eta$ Car prevented a perfect subtraction and so residuals
appear in the final image in frames containing $\eta$ Car. The continuum
images were then combined into a mosaic sequence. A coordinate axis was
added using a corresponding image obtained with the Digitised Sky Survey
(see see http://skyview.gsfc.nasa.gov/cgi-bin/surv\_comp.pl?dss) and the
program {\sc koords} \cite{Gooch96}. The same mosaic sequence and
coordinate axis were then applied to the emission-line images. Observations
of the spectroscopic standards BS 4013 and HD 105116 were used for the flux
calibration and continuum scaling factors. The central wavelength measured
varies by up to $\sim 10$ \kms\ between pixels ($\sim 1/7$ of the
FWHM). This can be corrected for by the use of a calibration lamp at
closely-spaced etalon settings. Normally line images are then obtained by
measuring at several plate spacings (typically 3--5) and fitting the
instrumental profile to the line.  However, because of the considerable
overheads this incurs, this was not attempted for the mapping mode of
operation used here. We estimate this introduces an uncertainty of $\sim
30$ per cent in the line fluxes. The seeing throughout the observations was
1.5 arcsec which, combined with tracking errors and co-addition of frames,
produced a positional uncertainty of $\pm$ 2.5 arcsec. The final image has
a $1 \sigma$ rms of $1 \times10^{-16}$ erg s$^{-1}$ cm$^{-2}$ pixel$^{-1}$.

\subsection{SPIREX/Abu observations}
\label{spirex-obs}
SPIREX is a 60-cm telescope commissioned in 1994 for operation at the
South Pole \cite{Herald90}. It was equipped in 1998 with the NOAO Abu IR
camera \cite{Fowler98}, which incorporates an engineering grade
$1024 \times 1024$ `Aladdin' InSb detector array able to image a circular
field of view of diameter 10.2 arcmin with a 0.6 arcsec pixel scale.

Observations were carried out on 1998 November 11 (during daylight) using
the narrow-band PAH filter. The response curve of this filter covers 3.262 -- 3.336 $\mu$m and is centred on 3.299 $\mu$m, with a half-power width of
0.074 $\mu$m. An integration time of 30 s was used for each frame. The
overlapping region from a total of 52 frames was used to obtain a final
image of 9 arcmin in diameter. The observing sequence consisted of 5 sky
frames followed by 10 object frames. All frames were dark subtracted and
flat-fielded. For the sky subtraction, a scaled median of the 6 nearest
frames, regardless of their actual image type, were subtracted from each
object image. The final image contains both PAH emission and continuum
emission. A coordinate axis was added using a corresponding image obtained
with the Digitised Sky Survey and the program {\sc koords}.

Observations of the standard star, $\delta$ Dor, observed 1998 November 19,
were used for flux calibration (L band magnitude of 3.711 mag). We estimate
the PAH fluxes to have an uncertainty of $\pm 5$ per cent. The diffraction
limit at \mbox{3.3 $\mu$m} of the telescope is 1.4 arcsec, comparable to
the typical ice-level seeing of the site. However, a combination of
tracking errors, tower shake and co-addition of frames limited the pointing
accuracy to $\pm$ 2.8 arcsec. The final image has a $1 \sigma$ rms noise of
\mbox{$2 \times 10^{-15}$ erg s$^{-1}$ cm$^{-2}$ pixel$^{-1}$}. 

The observed wavelength range includes the Pf $\delta$ hydrogen
recombination line. From the peak Br $\gamma$ emission flux measured
towards the southern edge of clump A1 ($6 \times10^{-15}$ erg s$^{-1}$
cm$^{-2}$ arcsec$^{-2}$) and the calculated relative intensities of the Pf
$\delta$ and Br $\gamma$ emission lines (0.26 for the case of N$_e=10^4$
cm$^{-3}$ and T$_e=10^4$ K, as listed in Table 6 of \citeNP{Hummer87}), we
estimate the emission from Pf $\delta$ is less than $2 \times 10^{-15}$ erg
s$^{-1}$ cm$^{-2}$ arcsec$^{-2}$. This value is the same as the $1 \sigma$
rms uncertainty of the final 3.29 $\mu$m image and therefore we assume
that the contribution from the Pf $\delta$ line is negligible.

\section{Results}

Fig.~\ref{multiwavelength}(b) shows a map of the H$_2$ emission towards the
Keyhole Nebula. The emission is distributed into seven discrete clumps of
less than 1 arcmin in size spread over an area of $4 \times 4$
arcmin$^2$. Each emission feature corresponds to a \CO emission clump
identified by \citeN{Cox951} and has been labelled accordingly, with the
exception of the faint H$_2$ emission clump located in the north-eastern
edge of the image which we have labelled U1.  No CO observations have been
made towards this clump. Fig.~\ref{multiwavelength}(c) shows an image of
the PAH emission over the same region of sky. The emission is also
distributed into clumps, all of which (except clump U1) can be matched
almost exactly with H$_2$ emission clumps. This confirms the H$_2$ emission
is caused by UV fluorescence and that the clumps are surrounded by
PDRs. Table~\ref{h2pah-table} lists the peak position, peak intensity,
total integrated intensity and size of each H$_2$ emission clump as well as
the corresponding PAH peak intensity and total integrated intensity.

\begin{table*}
\begin{minipage}{\textwidth}
\caption{\label{h2pah-table}Measured H$_2$ 1--0 S(1) and PAH emission
parameters of the clumps shown in Fig~\ref{multiwavelength}.}
\begin{tabular}{@{}ccccccccc}
\hline
Clump	& \multicolumn{2}{c}{H$_2$ Peak position}
        & H$_2$ Peak \footnote{H$_2$ image has a 1 $\sigma$ rms of $0.1
        \times 10^{-15}$ erg s$^{-1}$ cm$^{-2}$ arcsec$^{-2}$}
	& PAH Peak \footnote{PAH image has a $1 \sigma$ rms of $3 \times 10^{-15}$ erg s$^{-1}$ cm$^{-2}$ arcsec$^{-2}$}
	& Size
        & H$_2$ Total      
        & PAH Total 
        & PAH/H$_2$\\
	
	&
	& 
	& Intensity
	& Intensity
	&
	& Intensity
	& Intensity
	& Ratio\\
	
	& RA             
	& DEC
        & ($\times 10^{-15}$ erg s$^{-1}$
        & ($\times 10^{-14}$ erg s$^{-1}$
	& \arcsec\ $\times$ \arcsec
	& ($\times 10^{-13}$ erg
        & ($\times 10^{-12}$ erg
        & \\

	& (1950)
	& (1950)
	& cm$^{-2}$ arcsec$^{-2}$)
	& cm$^{-2}$ arcsec$^{-2}$)
	&
	& s$^{-1}$ cm$^{-2}$)
	& s$^{-1}$ cm$^{-2}$)  &\\

\hline
A1 &10 43 01 &$-59$ 22 05  &2.1 & 5.8 &$60 \times 60$ & 9.9 & 33    & 33\\
A2 &10 42 54 &$-59$ 22 09  &2.7 & 8.1 &$20 \times 40$ & 5.5 & 24    & 44\\
B  &10 43 11 &$-59$ 23 20  &1.8 & 4.7 &$15 \times 15$ & 1.8 & 3.6   & 20\\
C  &10 42 54 &$-59$ 25 06  &2.8 & 6.6 &$15 \times 25$ & 4.4 & 11   & 25\\
4  &10 42 35 &$-59$ 23 42  &2.2 & 2.8 &$15 \times 3$  & 3.7 & 9.1   & 25\\
5  &10 42 43 &$-59$ 22 06  &2.5 & 5.3 &$25 \times 7$  & 2.3 & 7.5   & 33\\
U1 &10 43 14 &$-59$ 21 53  &1.7 & $ < 0.8^{\it c}$ & $10 \times 10$ & 0.7 &
$< 0.1$\footnote{3$\sigma$ upperlimit}& $< 1$\\
\hline

\end{tabular} 
\end{minipage}
\end{table*}

\begin{figure*}
\centering \mbox{\subfigure[Clumps A1 (the `Kangaroo' nebula) and
A2.]{\psfig{file=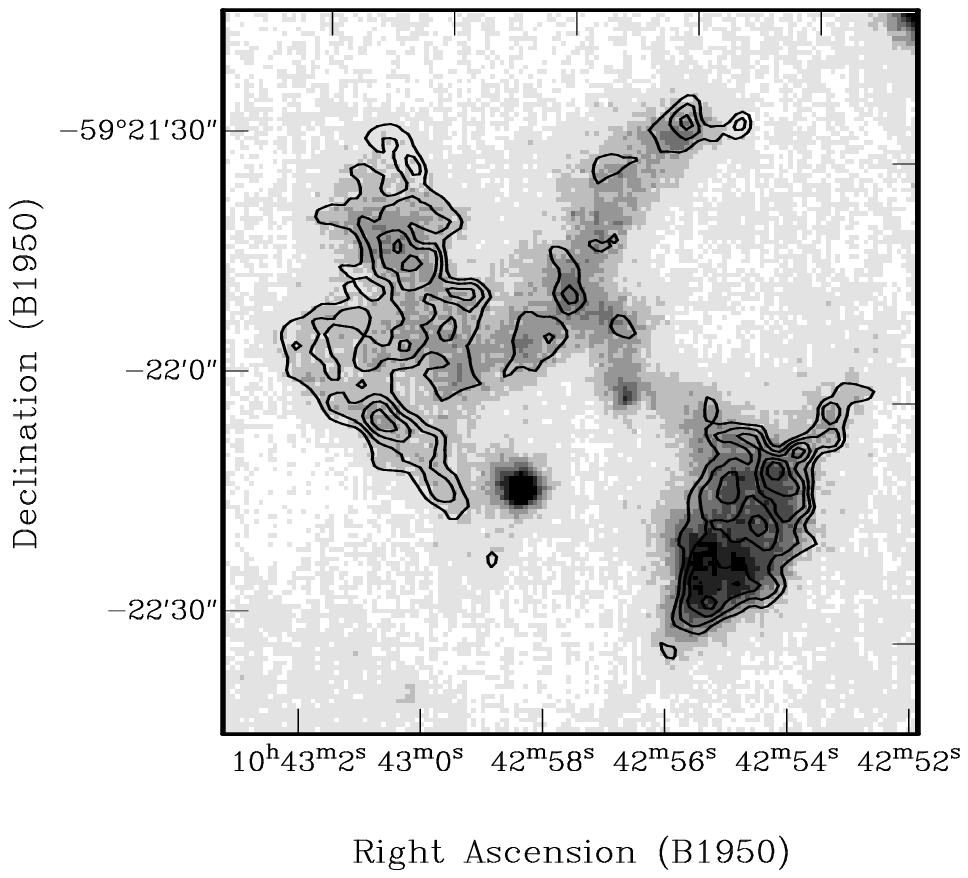,clip=,width=0.45\textwidth}}\quad
\subfigure[Clump 5]{\psfig{file=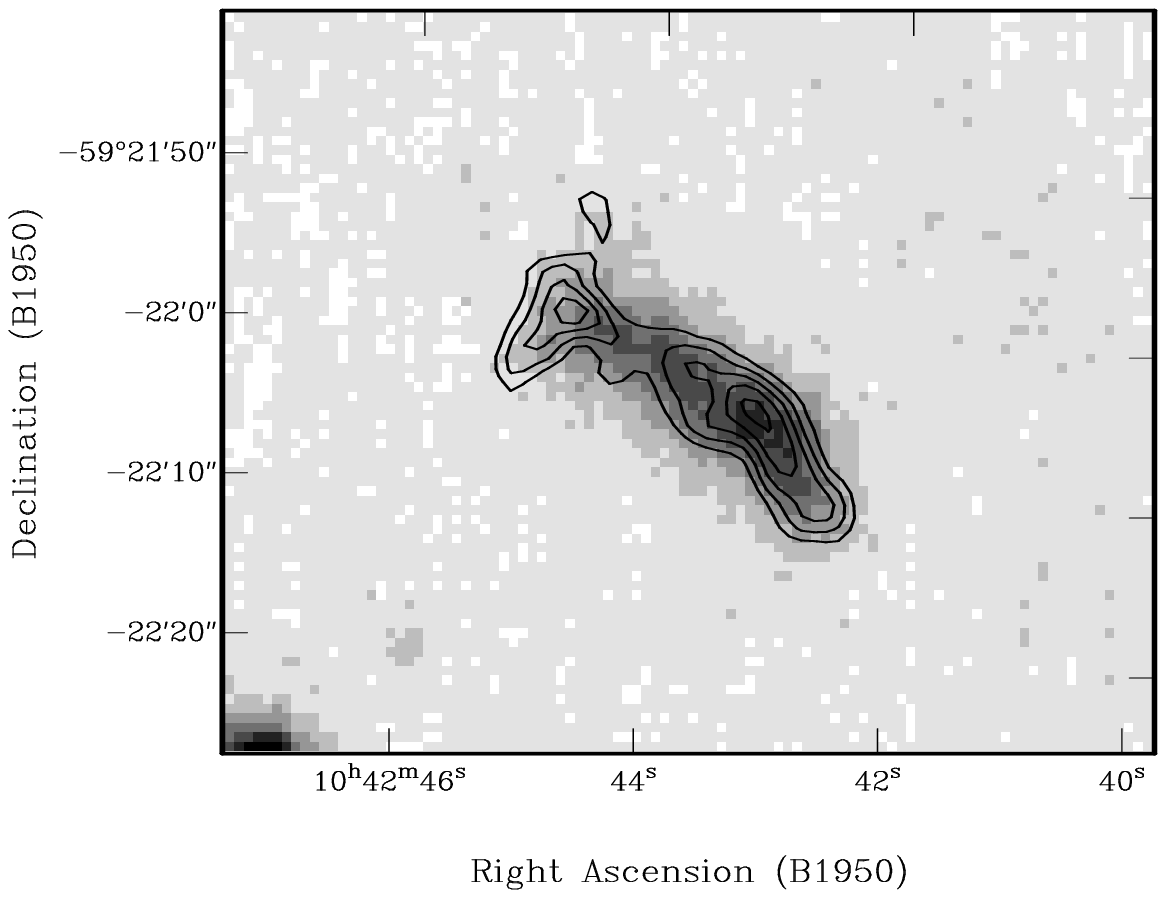,clip=,width=0.45\textwidth}}}
\mbox{\subfigure[Clump 4]{\psfig{file=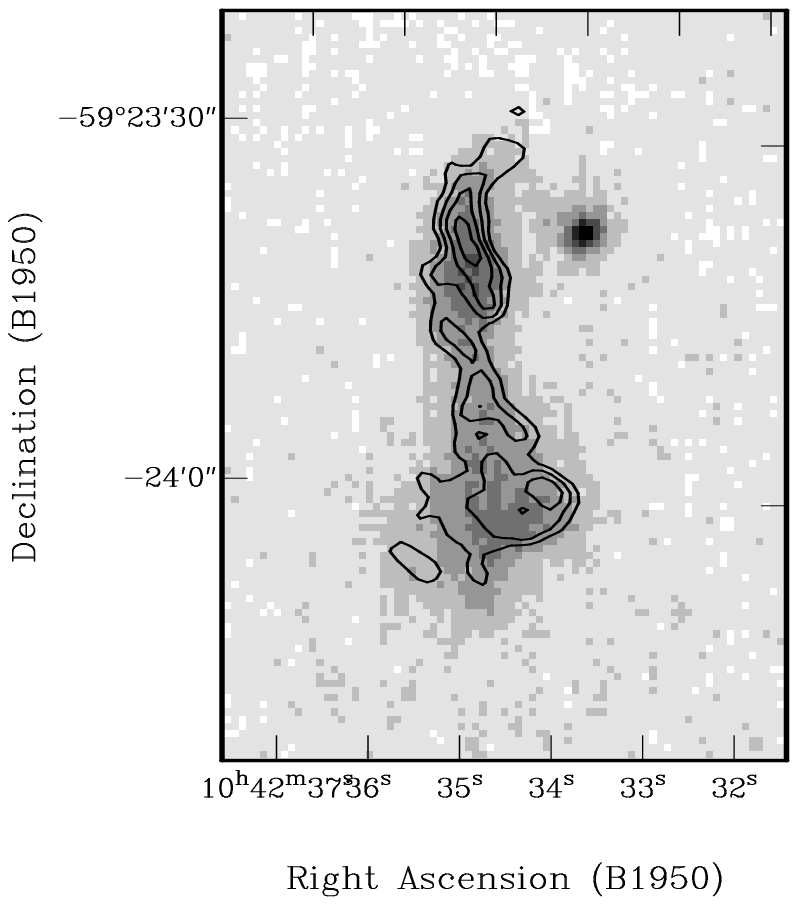,clip=,width=0.3\textwidth}}\quad
\subfigure[Clump B]{\psfig{file=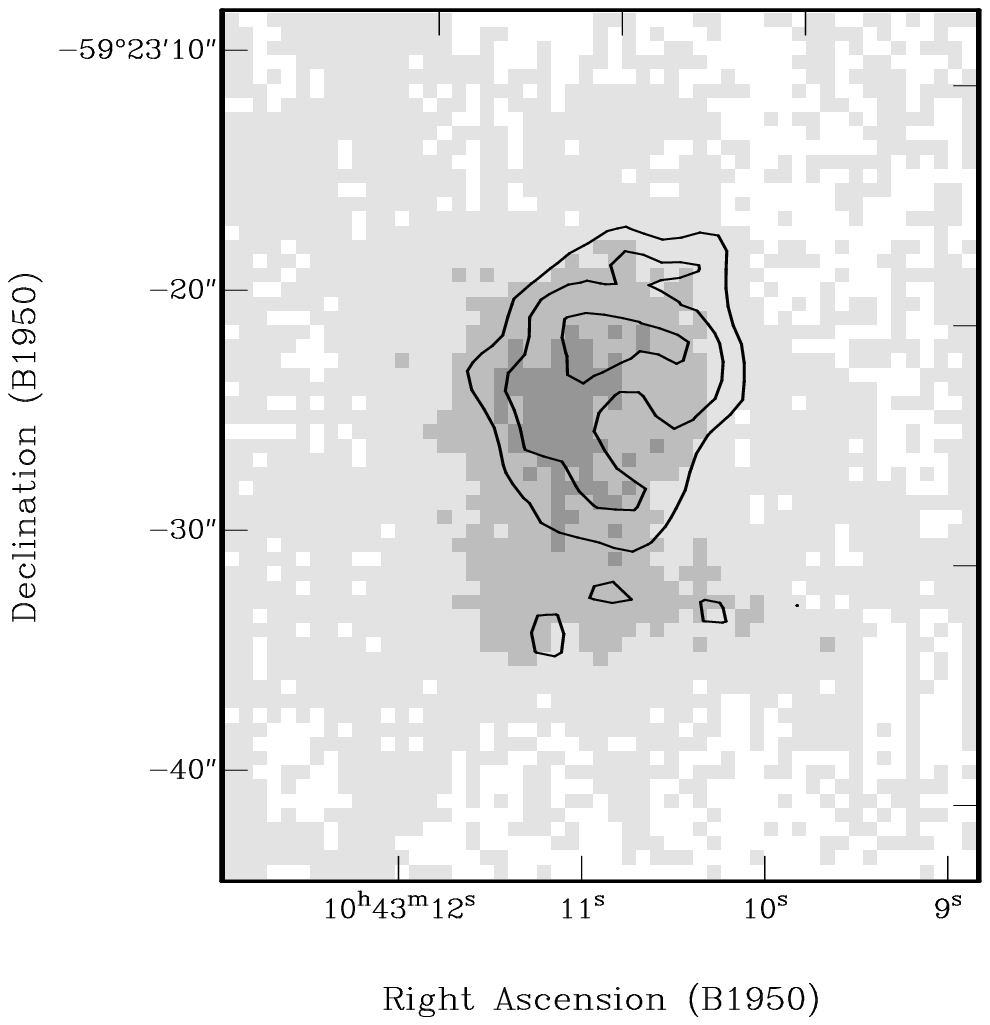,clip=,width=0.3\textwidth}}\quad
\subfigure[Clump C]{\psfig{file=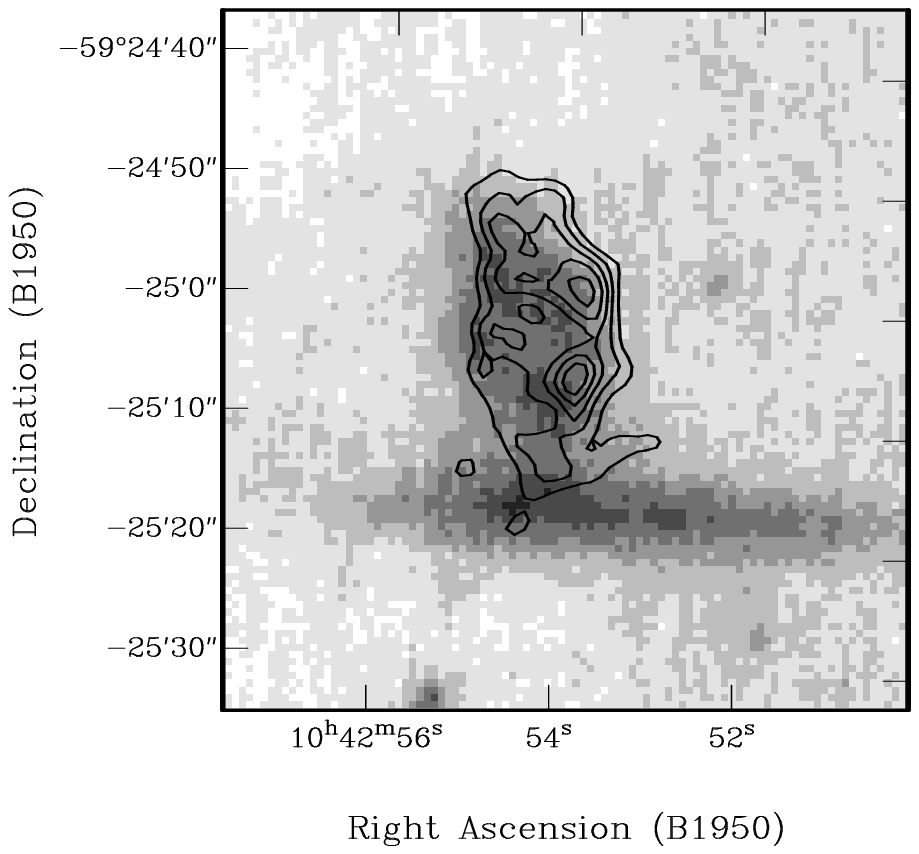,clip=,width=0.3\textwidth}}}
\caption{\label{h2pah-figure}PAH emission in grey-scale overlaid with
contours of the H$_2$ 1--0 S(1) line emission for Clumps A1, A2, 5, 4, B
and C. The contour levels are 5, 8, 12, 15, and $19 \times 10^{-16}$ erg
cm$^{-2}$ s$^{-1}$ arcsec$^{-2}$. Note the sharp line below Clump C is part
of the diffraction spike from $\eta$ Car.}
\end{figure*}

Fig.~\ref{h2pah-figure} shows the PAH emission in grey-scale overlaid with
contours of H$_2$ emission for clumps A1\footnote{We have dubbed clump A1
the `Kangaroo Nebula' owing to its morphological resemblance to an Antipodean
marsupial, see Ryder et al. (1998).}, A2, 4, 5, B and C . The overall
distribution of the H$_2$ and PAH emission is similar for all the clumps,
however, there are small-scale differences. The emission is not uniform but
exhibits many small bright knots and in some cases the H$_2$ emission knots
do not correspond to PAH emission knots. This is to be expected considering
that the H$_2$ and PAH emission zones within a PDR may not completely
coincide with one another. The bright H$_2$ and PAH emission knots are
pinpointing high-density regions at different UV optical depths within the
molecular clumps.

The ratio of the PAH total intensity to the H$_2$ 1--0 S(1) total intensity
is reasonably consistent from clump to clump. Clump A2 has the highest
ratio (44), while clump B has the lowest (20).

\section{Discussion}
\subsection{Geometry}
\begin{figure}
\psfig{file=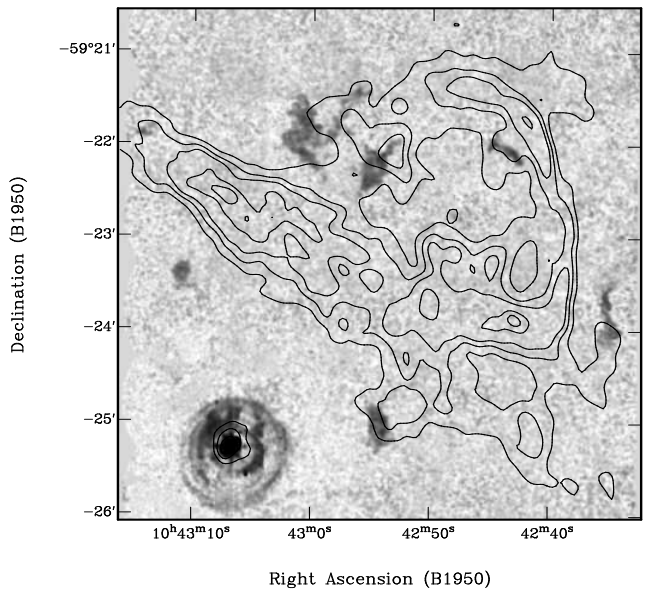,silent=,width=\columnwidth}
\caption{\label{h2-cont}H$_2$ emission (grey scale) overlaid with contours
of 4.8-GHz continuum emission taken from Brooks et al. (in
preparation). The contour levels are 0.02, 0.03, 0.04, 0.05, 0.06, 0.07 Jy
beam$^{-1}$. Note the contours and emission circles towards $\eta$ Car are
artifacts.}
\end{figure}

Fig.~\ref{h2-cont} shows the grey-scale H$_2$ emission superimposed with
the 4.8-GHz radio continuum emission from Brooks et al. (in
preparation). The H$_2$ clumps are distributed around the edges of the
continuum emission. The absence of the Keyhole structure in the radio
continuum emission indicates that the Keyhole is in the foreground of the
nebula and obscuring it.
 
When comparing the H$_2$ and PAH emission with the optical emission of
Fig.~\ref{multiwavelength}(a) it appears that all the emission clumps
correspond to dark optical features. Clumps A1, A2, 4 and 5 correspond to
prominent patches which have bright rims, while clumps B and C coincide
with very faint, filamentary patches. It is worth noting that no H$_2$ or
PAH emission was detected towards the most prominent optical
obscuration---the Keyhole itself, or towards the boomerang-shaped filament
to the western edge of the image. Neither of these latter two dark features
appears to have a bright rim, unlike clumps A1, A2, 4 and 5.

The Keyhole has been traced in \CO emission by \citeN{Cox951} and
corresponds to clumps 1, 2, and 3 (see Table~\ref{co-clumps}). These three
clumps are the only molecular clumps detected by \citeN{Cox951} which show
no H$_2$ or PAH emission.

\begin{figure}
\psfig{file=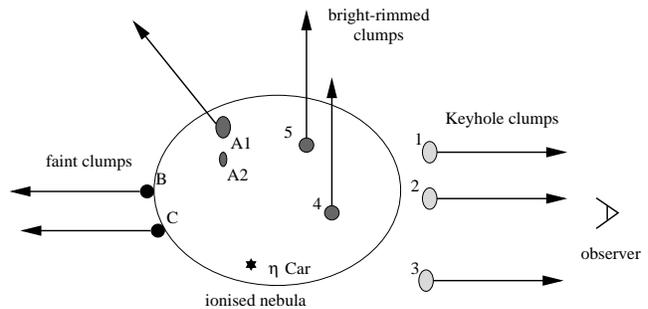,silent=,width=\columnwidth}
\caption{\label{schematic-results}A schematic representation of the three
dimensional geometry of the region, based on the data presented here. The
region of ionized gas is bounded by an ellipse and the star denotes the
location of $\eta$ Car. Each clump, with the exception of U1, has been
grouped according to its velocity and optical features (see
Table~\ref{co-clumps}). The arrows indicate the direction in which the
clumps are moving with respect to the ionized gas.}
\end{figure}

As listed in Table~\ref{co-clumps}, the velocities with respect to the
local stand of rest (LSR) of the molecular clumps taken from their \CO
emission vary from $-32$ to \mbox{$-8$ \kms}. When comparing these
velocities with the clump's corresponding optical, PAH and H$_2$ emission
features there appears to be an interesting trend. Firstly, the clumps with
the most negative velocities (clumps 1, 2, and 3) are those which have the
most prominent optical obscuration and also are the only clumps with no
associated PAH and H$_2$ emission. Secondly, those clumps with intermediate
velocities (clumps A1, A2, 4 and 5) correspond to the optical features with
bright rims. Thirdly, the clumps which have the least negative velocities
(clumps B and C) correspond to the very faint patches. Using these results
we present a model for the Keyhole Nebula whereby the velocities of the
clumps are associated with positions along our line of sight; clumps 1, 2,
and 3 (the Keyhole) are in front of the nebula, clumps A1, A2, 4 and 5 are
in it and clumps B and C are located at the far side. This model is
represented schematically in Fig.~\ref{schematic-results}.

Radio recombination-line observations of the ionized gas in this region
(Car II) reveal a complex velocity structure which has been interpreted as
an expanding shell centred close to $\eta$ Car (\citeNP{Huchtmeier75},
\citeNP{Deharveng75}). There are two main velocity components whose
relative fluxes vary across the region; one having LSR velocities between $-6$ to $-17$ \kms; and the other having LSR velocities between $-23$ to $-41$
\kms. These velocities are consistent with clumps A1, A2, 4 and 5
(i.e. those with bright rims), which have LSR velocities between $-17$ and
$-25$ \kms, being located in the ionized region and the remaining clumps,
with LSR velocities near $-8$ and $-30$ \kms, being situated at the far and
near side, respectively. The systematic LSR velocity of $\eta$ Car is taken
to be $-18.5 \pm 10$ \kms, similar to the other O-type stars in the region
\cite{Davidson972}, placing it in the middle of the range.

\subsection{Dynamics}

It is generally thought that the strong stellar winds from the massive
stars belonging to Tr 16, in particular $\eta$ Car, are responsible for the
complex dynamics of the ionized and molecular gas in the Keyhole region
(e.g. \citeNP{Cox951}, \citeNP{Cox952} and \citeNP{Deharveng75}). Here we
extend the analysis of previous studies to include a better estimate of the
mechanical luminosity associated with $\eta$ Car and the energy conversion
efficiency factor required to sustain the kinetic energy of a typical
clump.

The kinetic energy of a typical clump, K$_{\mathrm C}$, with a mass of 10
\Msun\ and a velocity of 10 \kms, is of the order 10$^{46}$ erg. We can
equate this value with the mechanical luminosity, L$_{\mathrm W}$, of the
stellar wind of $\eta$ Car using, \mbox{K$_{\mathrm C}$ $\sim$ L$_{\mathrm
W}$\,$\Omega\,\,\tau\,\epsilon$}, in order to obtain the required
conversion efficiency factor, $\epsilon$. We estimate the solid angle,
$\Omega$, of a typical clump of cross-section 1 arcmin located 3.5 arcmin
from $\eta$ Car is $\sim 0.06$. The mechanical luminosity of the wind for
$\eta$ Car, using a mass-loss rate of $5 \times 10^{-4}$ \Msun\ yr$^{-1}$
\cite{Davidson97} and a wind velocity of 700 \kms\ \cite{Lamers89} is
\mbox{$\sim 8 \times 10^{37}$ erg}. If we take the time period, $\tau$,
to be equal to the passage time across the ionized nebula, $\sim 10^{5}$
years (2 arcmin at 10 \kms), then the efficiency factor need only be $\sim
0.01$ per cent.

In practice $\tau$ is likely to be very much less than 10$^{5}$ years, as
the clumps will rapidly decelerate as they sweep up material.  We estimate
its value following \citeN{Dyson97}. They show that the radius of a stellar
wind-blown bubble is given by $\sim$ (L$_{\mathrm W}/n \tau^{3/5}$. For a
typical clump-mass of 10 \Msun, the required density ($n$) of the swept-up
gas is $\sim 1$ 000 cm$^{-3}$. This yields $\tau \sim 10^{3}$ years for the
$\eta$ Car wind which leads to a required efficiency factor of $\sim 1$ per
cent. This efficiency factor is still much less than the 20 per cent
efficiency factor estimated by \citeN{Dyson97} for this process for a
steady wind.

Thus, the kinetic energy of the clumps can readily be explained by the
mechanical energy of a wind from $\eta$ Car. We conclude that the clumps
represent swept-up ambient cloud material which is being driven out of the
ionized region by such a wind. In this scenario we would expect the
swept-up mass of each clump to be much greater than its wind mass (given by
$\frac{\dot{M}\,\Omega\,\tau}{4\,\pi}$). Substituting the figures above, we
obtain 0.003 \Msun\ for the wind mass, considerably less than the typical
swept-up mass of 10 \Msun, as expected.

\subsection{Excitation}

The morphology of the H$_2$ and PAH emission strongly suggests that UV
fluorescence is responsible for the excitation. This conclusion can be
quantitatively tested by comparing the observed fluxes with those expected
for the physical conditions within the nebula.

We first estimate the ionizing ({\it hv} = 13.6 eV $\rightarrow$ $\infty$)
and far-UV ({\it hv} = 6 eV $\rightarrow$ 13.6 eV) fluxes incident on clump
A1 (taken to be 1 arcmin across), from both $\eta$ Car (3.5 arcmin away)
and the O3 star HD 303308 (2.8 arcmin away). HD 303308 is taken to have
T$_{\mathrm eff}$ $\sim$ 47 500 K and radius $ 1\times 10^{12}$ cm,
following \citeN{Panagia78}. Using Kurucz's (1997) model stellar spectra we
find F$_{\mathrm UV}$ = 0.34 erg s$^{-1}$ cm$^{-2}$ and \mbox{F$_{\mathrm
FUV}$ = 0.26 erg s$^{-1}$ cm$^{-2}$}. The spectral type of $\eta$ Car is
uncertain. Based on the observations of \citeN{Cox953} at wavelengths
longer than 1 $\mu $m, its luminosity has been estimated by
\citeN{Davidson97} to be $5 \times 10^{6}$ \Lsun. At such high
luminosities, blackbody and Kurucz models indicate that we can assume the
bulk of the flux is distributed over the ionizing and far-UV wavelengths
and that F$_{\mathrm UV}$ $\sim$ F$_{\mathrm FUV}$. Using a stellar radius
of $6\times10^{12}$ cm \cite{Davidson97} we find that for $\eta$ Car
F$_{\mathrm UV}$ $\sim$ F$_{\mathrm FUV}$ $\sim$ 13 erg s$^{-1}$ cm$^{-2}$,
which is considerably greater than the corresponding fluxes from HD 303308.

We thus estimate the total number of ionizing UV photons incident on clump
A1 is N$_{\mathrm Lycont} \sim 3 \times10^{48}$ photons s$^{-1}$ and
are mostly from $\eta$ Car. This number is an upper limit, and may be
reduced if either internal extinction is present or the distance from the
source is greater than its projected value. We have obtained an additional
observation of Br $\gamma$ emission at \mbox{2.166 $\mu$m} using
UNSWIRF. The total Br $\gamma$ flux from a region covering 0.7
arcmin$^{2}$, adjacent to the northwest-southeast edge of clump A1, is $3.9
\times 10^{-11}$ erg s$^{-1}$ cm$^{-2}$. From this, assuming that each
Lyman-continuum photon produces 70 Br $\gamma$ photons, we derive
N$_{\mathrm Lycont} \sim 2\times 10^{48}$ photons s$^{-1}$, remarkably
similar to our estimate based on the stellar luminosities.

We estimate the total far-UV radiation field on clump A1 is F$_{\mathrm
FUV} \sim  13$ erg s$^{-1}$ cm$^{-2}$, which is equivalent to $\sim 8$
000G$_{\mathrm 0}$, where G$_{\mathrm 0} = 1.6 \times 10^{-3}$ erg s$^{-1}$ cm$^{-2}$ and is the average interstellar radiation field
\cite{Habing68}. Using a far-UV radiation field of $10^{4}$ G$_{\mathrm 0}$
and density of 10$^{4}$ cm$^{-3}$, PDR models (e.g. \citeNP{Burton90})
predict an H$_2$ 1-0 S(1) line flux of \mbox{$1.4 \times 10^{-5}$ erg
s$^{-1}$ cm$^{-2}$ sr$^{-1}$}, which, integrated over a 1 arcmin region,
gives a total line flux of $1.2 \times 10^{-12}$ erg s$^{-1}$
cm$^{-2}$. This is remarkably similar to that measured from clump A1,
$9.9 \times 10^{-13}$ erg s$^{-1}$ cm$^{-2}$
(Table~\ref{h2pah-table}). Furthermore, if the density were $10^{3}$
cm$^{-3}$ the prediction would be a factor of 10 smaller, while if the
density were 10$^{5}$ cm$^{-3}$ it would be a factor of 5 higher. A density
of 10$^{4}$ cm$^{-3}$ is consistent with CO and CS observations towards
clump A1 by \citeN{Cox951}, which suggest a value in excess of $10^{3}$
cm$^{-3}$.

A typical clump with a surface area of 1 arcmin$^2$, hydrogen molecule
density of $10^{4}$ cm$^{-3}$ and mass of \mbox{10 \Msun} yields an
emitting thickness, A$_{v}$ $\sim$ 1 (in a direction away from the exciting
star) of $\sim 10^{17}$ cm (4 arcsec at 2.2 kpc). This order of magnitude
calculation implies widths similar to those observed for the elongated
clumps, 4 and 5, suggesting that they are seen edge-on. These two clumps
and the larger clumps, A1 and A2, are thought to be situated inside the
\HII\ region (see Fig. 4). Clumps 4 and 5 may be moving edge on, in the
plane of the sky, and clumps A1 and A2 may be moving mostly face on, away
from us.

The emission measure, {\it n}$_{\mathrm \it e}^{2}$ L, derived from the Br
$\gamma$ line flux is $6 \times 10^{5}$ cm$^{-6}$ pc. Taking a typical
clump thickness as representative of L gives L $\sim  4$ arcsec (or 0.03
parsec based on the above estimate). This yields an electron density, {\it
n$_{\mathrm e}$} $\sim 4 \times 10^{3}$ cm$^{-3}$ for the ionized gas,
which is intermediate in value between the estimates for the initial
density and that swept-up in the clumps. The emission measure is also
consistent with the value of $2 \times 10^5$ cm$^{-6}$ pc obtained from
measurements of the 4.8 GHz continuum emission over the same region (Brooks
et al., in preparation).

The H$_2$ and Br $\gamma$ line emission are thus consistent with UV
excitation, from $\eta$ Car. The clumps themselves are swept-up cloud
material, likely driven by the mechanical luminosity of a strong wind from
$\eta$ Car. This wind is strong enough to also produce shocked H$_2$
emission. Since there is now no clear evidence for any such emission this
wind cannot still be interacting with the clumps. The clumps must currently
be coasting while being externally heated by the radiation field from the
nearby stars. Considering the episodic events of $\eta$ Car's evolution the
current state of the clumps may only represent a quiescent stage in their
life and they may be overrun in the future by another fast stellar wind
from $\eta$ Car, resulting in shocked emission. This type of variability
however would be on time scales of the order of 1 000 years, considerably
longer than the observed 5.54-year brightness fluctuation cycle of $\eta$
Car \cite{Daminelli97}.

It is possible to estimate the number of PAH molecules that are
emitting from the surfaces of the clumps, using their 3.29 $\mu$m flux and
the incident far-UV radiation field. Following Fig. 13 of
\citeN{Allamandola89}, the typical energy emitted in the 3.3 $\mu$m band by
a PAH molecule, excited via far-UV fluorescence to 30 000 cm$^{-1}$ of
vibrational energy, is $\sim 6 \times 10^{-14}$ erg s$^{-1}$
molecule$^{-1}$ sr$^{-1}$. For an incident far-UV field, G$_0 \sim 10^4$,
there are $\sim 10^{12}$ photons cm$^{-2}$ s$^{-1}$. Taking a typical
absorption cross-section for a PAH molecule of $7 \times 10^{-18}$ cm$^2$
per C atom (Tielens, private communication), then for a typical 50-atom
PAH molecule we expect an excitation rate of $\sim 4 \times 10^{-4}$
s$^{-1}$, or about one excitation each hour.  This energy is rapidly
re-emitted through infrared fluorescence in the various PAH emission bands,
including the 3.3 $\mu$m band. For the peak 3.29 $\mu$m flux of $3 \times
10^{-3}$ erg s$^{-1}$ cm$^{-2}$ sr$^{-1}$ measured from clump A1, we thus
derive an emitting PAH column density of $\sim 10^{14}$ cm$^{-2}$ from this
region. These estimates show that a significant fraction of the C-atoms may
reside in PAH molecules. For instance, if the PAH emitting region were
confined to an optical depth A$_{v}$ $<$ 0.1 from the front surface of the
cloud, it would imply a C-abundance in the PAH molecules of $\sim 3 \times
10^{-5}$ relative to hydrogen, or around 30 per cent of the total carbon,
assuming typical interstellar carbon abundances.

If the rate of destruction of H$_2$ molecules by far-UV radiation exceeds
their formation rate on dust grains, it is possible that photodissociation
equilibrium will not exist in the molecular clumps. For instance,
\citeN{Hollenbach95} derive a characteristic time-scale of
$5\times10^8n^{-1}$ years to establish equilibrium. This translates to
$\sim5\times10^4$ years using the densities derived here, which is greater
than the assumed age of $\sim10^3$ years for the wind. If this were the
case the ionization-dissociation front impinging on the clumps will
constantly be exposing fresh molecular material to the radiation field and
the fluorescent line intensities will be higher than steady-state model
predictions. However the front, moving at perhaps $\sim1$ \kms, would not
have had time to propagate through the clumps. If photodissociation
equilibrium does not exist we would expect the ratio of the $\nu$=1--0/2--1
S(1) lines to preferentially show pure fluorescent values ($\sim2$) where
the  $\nu$=1--0 S(1) line is strongest (see \citeNP{Allen99}). This
could be tested with further observations.

\section{Conclusions}

We have imaged the Keyhole region of the Carina Nebula in the H$_2$ 1-0
S(1) line at 2.122 $\mu$m and in the PAH emission feature at 3.29 $\mu$m,
and compared the results to optical, radio continuum and \CO images of the
region. The H$_2$ and PAH images are remarkably similar, indicating they
are fluorescently excited, and show a series of emission clumps typically
30--60 arcsec in extent distributed around the edges of the radio
emission. When compared to the optical and CO images three categories of
clumps are apparent: (i) dark regions of optical obscuration with CO but no
H$_2$ or PAH emission (clumps 1, 2, and 3 which form the Keyhole feature),
(ii) faint, optically dark clumps with bright rims and H$_2$ and PAH
emission (clumps A1, A2, 4 and 5) and (iii) faint optically dark clumps
with H$_2$ and PAH emission (clumps B and C).

The clumps are typically of mass \mbox{10 \Msun} and appear to have been
swept up from ambient cloud material by a wind from $\eta$ Car. They have
been driven in all directions away from $\eta$ Car, and must now be
coasting at speeds of \mbox{$\sim$ 10 \kms} as there is no evidence of any
shocked emission. The prominent dark clumps of the Keyhole are coming
towards us and are outside the \HII\ region. The bright-rimmed clumps A1,
A2, 4 and 5 are within the \HII\ region and thus are being heated by the
radiation field. Clumps such as A1 are seen face-on whereas clumps 4 and 5
are seen edge-on. Finally the faint dark clumps B and C are on the far side
of the \HII\ region and have been driven away from $\eta$ Car in the
opposite direction to ourselves. The clumps comprise the last remnants of
the ambient molecular cloud from which $\eta$ Car formed.

\section{Acknowledgements}

We would like to thank Lori Allen, Stuart Ryder and John Whiteoak for their
help during the observations, and Rodney Marks and members of CARA (the
Center for Astrophysical Research in Antarctica), in particular Al Fowler
and Al Harper, for obtaining the SPIREX/Abu data. We thank David Malin for
providing the colour image of the Keyhole Nebula and Xander Tielens for
useful discussions concerning PAH abundances. We are grateful to Stuart
Ryder for the UNSWIRF reduction software. We also thank Pierre Cox for
helpful comments regarding the manuscript. KJB acknowledges the support of
an Australian Post-graduate Award. This work has been supported by a grant
from the Australian Research Council.

\end{document}